\documentclass{elsart}

\usepackage{natbib}

\usepackage{epsfig}

\usepackage{amssymb}

\begin{document}

\begin{frontmatter}


\title{Vortex nucleation and flux front propagation
in type II superconductors}
\author[roma]{Stefano Zapperi \corauthref{cor1}}
\ead{stefano.zapperi@roma1.infn.it}
\corauth[cor1]{}
\address[roma]{INFM, SMC and  UdR Roma 1, Dipartimento di Fisica, Universit\`a
``La Sapienza'', P.le A. Moro 2, 00185 Roma, Italy.}
\author[ufc]{Jos\'e S. Andrade Jr.}
\address[ufc]{Departamento de F\'{\i}sica, Universidade Federal do
Cear\'a, 60451-970 Fortaleza, Cear\'a, Brazil.}
\author[nw]{Andr\'e A. Moreira}
\address[nw]{Department of Chemical Engineering, Northwestern
University, Evanston, IL 60208, USA}

\begin{abstract}
We study flux penetration in a disordered type II superconductor by
simulations of interacting vortices, using a Monte Carlo method for
vortex nucleation. Our results show that a detailed description of the
nucleation process yields a correction to the scaling laws usually
associated with flux front invasion. We propose a simple model to
account for these corrections.
\end{abstract}


\end{frontmatter}

\section{Introduction}
The magnetization of type II is traditionally described in terms of
the Bean model \cite{bean}: magnetic flux enters into the sample from
the boundaries, forming a flux gradient that is pinned by disorder. At
the microscopic level, the process takes place through the nucleation
of vortex lines carrying each a flux quantum \cite{Blatter}. The Bean
model provides a phenomenological picture of average magnetization
properties, such as hysteresis and thermal relaxation \cite{kim}, but
it is inadequate to describe fluctuations, which turn out to be quite
important. As it has been shown experimentally using magnetoptical
methods, flux fronts are typically rough or even fractal
\cite{rough,fractal,fractal2}.

Recently, we have shown that the flux penetration in a disordered
superconductor can be described by a disordered non-linear diffusion
equation \cite{zms}. The equation can be obtained performing a
coarse-graining of the microscopic equation of motion of the vortices.
In the absence of pinning, it reduces to the model introduced in
Ref.~\cite{dorog}. This model has been solved analytically to provide
expressions for the dynamics of the front for different boundary
conditions \cite{dorog,gilc} and the results are in perfect agreement
with vortex dynamics simulations \cite{MOR-02}.  When quenched
disorder is included in the diffusion equations, flux fronts are
pinned in agreement with individual vortex simulations. Varying the
parameters of the equation, we observe a crossover from flat to
fractal flux fronts, consistent with experimental observations. The
value of the fractal dimension suggests that the strong disorder limit
is described by gradient percolation
\cite{zms}.

Here we reconsider the influence of boundary conditions in the front
dynamics. In Ref.~\cite{MOR-02} we have analyzed several types of
boundary conditions determining the way vortices enter into the sample
depending of the experimental setup one would like to model.  For
instance, a constant applied magnetic field can be approximated by a
constant vortex density at the boundary of the sample
\cite{dorog,gilc}.  This assumption simplifies the real vortex
nucleation process and allows for straightforward numerical and
analytical approaches \cite{zms,dorog,gilc,MOR-02}.  A more accurate
description of vortex nucleation can be obtained combining the vortex
dynamics simulations with the Monte Carlo method \cite{SHU-97}. Using
this method we find that the simple widely employed approximations for
the boundary conditions are only asymptotically true. At short times
the details of the nucleation process affects the expected scaling
behavior for the front dynamics.  We are able to quantitatively
estimate the corrections to scaling, using a simple front dynamics
model in the spirit of Washburn approach to fluid imbibition \cite{dube}

\section{Simulations}

We use here the model introduced in Ref.~\cite{SHU-97}, which combines
a typical vortex dynamics simulation scheme with a Monte Carlo method
for vortex nucleation.  In a very large sample with a constant
magnetic field $H$ oriented along the $z$ axis, vortices are modeled
as a set of interacting particles performing an overdamped motion in
the $xy$ plane.

The Gibbs potential associated to $N$ vortices of coordinates
$\vec{r}_i$ can be written as
\begin{equation}
G=\sum_{ij} E_{12}(\vec{r}_i-{r}_j)+\sum_i \tau_i(x_i)-\frac{H}{4\pi}\sum_i \phi_i(x_i),
\label{eq:gibbs}
\end{equation}
where in an infinite system
$E_{12}(\vec{r})=[\Phi_0^2/(8\pi^2\lambda^2)]K_0(|\vec{r}|/\lambda)$
is the vortex-vortex interaction in the London-London theory, with
flux quantum $\Phi_0$ and penetration length $\lambda$. Here we
consider a semi-infinite system, bounded by the $y=0$ line, and
consequently we add to the vortex-vortex interaction a term accounting
for the interaction between each vortex and the image of the others
\cite{SHU-97}.  The term $\tau_i(x_i)$, where $x_i$ is the distance
between the vortex $i$ and the sample surface, represents the
interaction between each vortex and its own image \cite{SHU-97}.
Finally the external magnetic field gives rise to a sort of chemical
potential with $\phi(x)\equiv \Phi_0 (1-\exp(-x/\lambda)$. In
Ref.~\cite{SHU-97} the interaction energy with random pinning centers
is also included to Eq.~\ref{eq:gibbs}, while here we restrict our
attention to a clean system.

The vortices evolve according to an overdamped equation of motion
$\Gamma \partial_t \vec{r}_i = -\partial_{r_i} G$, where $\Gamma$ is a
damping constant. The equation of motion is integrated numerically and
after each integration step a zero temperature Monte Carlo step is
performed: a new vortex is nucleated at a random position in a strip
of length $\lambda$ close to the sample boundary if the Gibbs
potential is reduced. In practice, we consider a system of size
$L_y=20\lambda$ and $L_x=100\lambda$, with periodic boundary
conditions along the $y$ direction. At the beginning of the
simulation, we start with an empty lattice and vortices are nucleated
close to the $x=0$ boundary. Since we are only interested in the
transient behavior the boundary condition at $x=L_x$ is inessential.

\begin{figure}
\epsfig{file=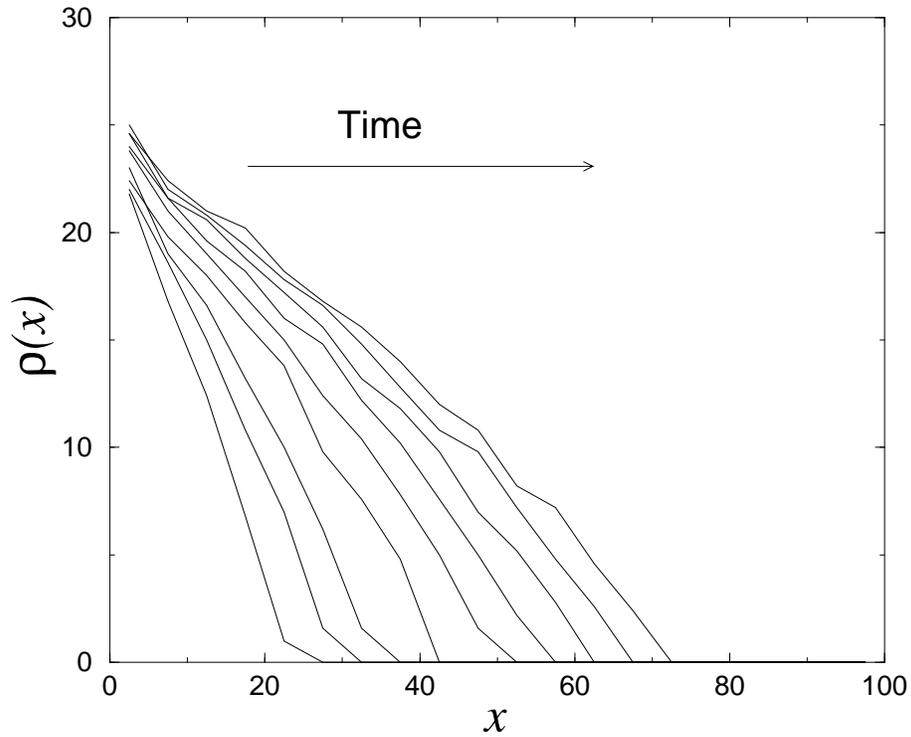,width=12cm}
\caption{The density profile at different times $t$ obtained from 
numerical simulations.}
\label{fig:1}
\end{figure}

\begin{figure}
\epsfig{file=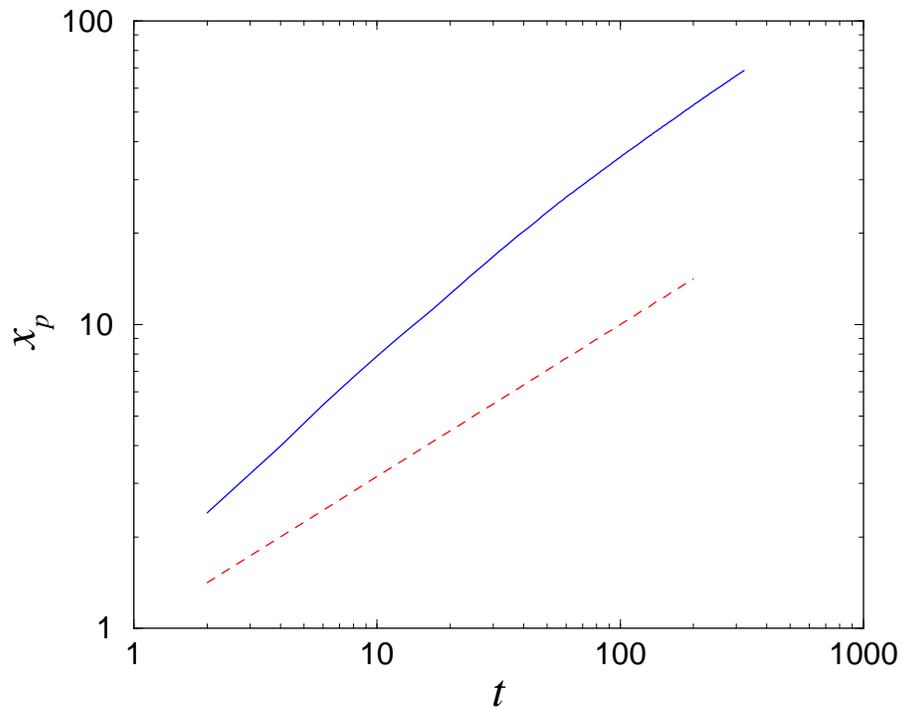,width=12cm}
\caption{The position of the front at different times $t$ obtained from 
numerical simulations. The dashed line is the $t^{1/2}$ prediction}
\label{fig:2}
\end{figure}

As the vortices are nucleated, they are pushed toward the interior of
the sample giving rise to a density profile. The evolution of the profile
is reported in Fig.~\ref{fig:1}. The results are in slight disagreement 
with the simplified boundary condition used in Refs.~\cite{dorog,gilc,MOR-02}:
a constant boundary vortex density.
Fig.~\ref{fig:1} clearly shows that the boundary density increases. 
In addition, Fig.~\ref{fig:2} shows that the front position does not
evolves as a power law, $x_p \sim t^{1/2}$, as predicted by the theory 
\cite{dorog,gilc,MOR-02}. 

\section{A simple model for front dynamics}

In order to account for the behavior observed in numerical
simulations, we consider a simple model for the front dynamics
\cite{MOR-02}.  The front is driven by the density gradient which can
be can be estimated simply as $\nabla \rho \simeq \rho_0/x_p$,
where $x_p$ is the front position and $\rho_0=\rho(0,t)$ is the boundary
density. Thus the equation of motion of the front can be written as $
\Gamma dx_p/dt = a \rho_0/x_p$, where $a$ is the interaction parameter
computed in Ref.~\cite{MOR-02}. In order to solve this equation, we
have to specify the dynamics of the boundary density, which to a first
approximation can be written as $\tau d\rho_0/dt = \rho_H-\rho_0$
where $\rho_H\propto H$ is the asymptotic value of the vortex density,
and $\tau$ a characteristic time.

Solving the two differential equations, we obtain
\begin{eqnarray}
\rho_0 &= &\rho_H (1-\exp(-t/\tau)) \label{eq:rho_0} \\
x_p(t)& = & \left(\frac{2a \rho_H}{\Gamma}\right)^{1/2} (t+\tau(1-\exp(-t/\tau)))^{1/2}.
\end{eqnarray}
In order to compare this result with numerical simulations, we plot on
the same graph the numerically calculated $\rho_0$ and $dx_p^{2}/dt$
together with the theoretical prediction from Eq.~\ref{eq:rho_0}. The
numerical result can be well fit by the model with $\tau \simeq 10$.

\begin{figure}
\epsfig{file=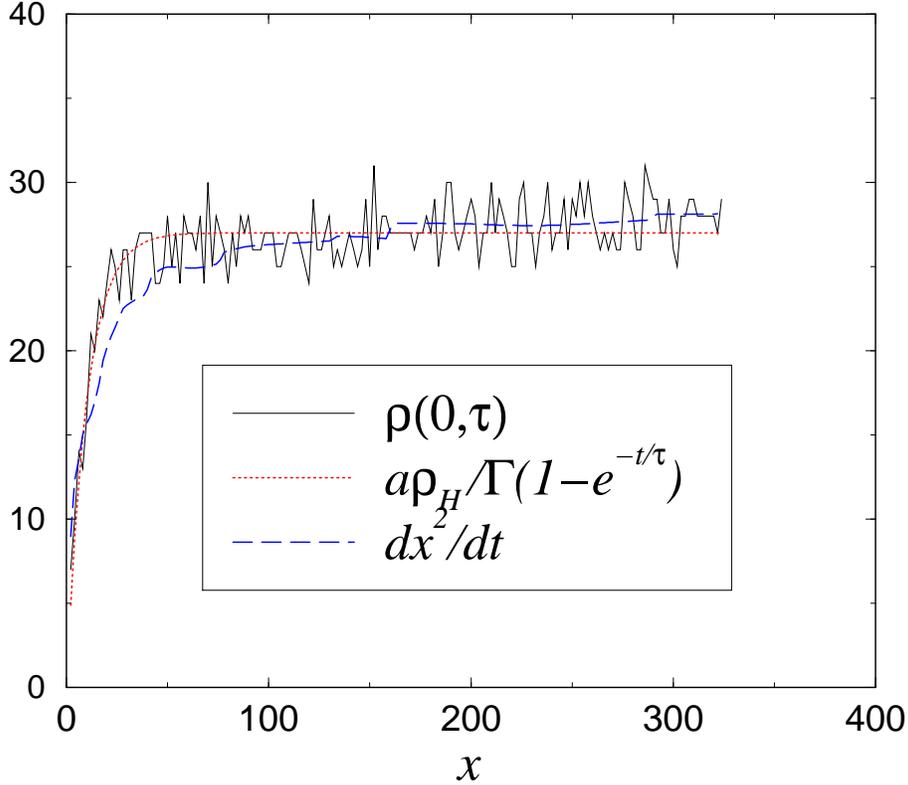,width=12cm}
\caption{A comparison between the front dynamics model
and the numerical simulations.}
\label{fig:3}
\end{figure}

\section{Conclusions}

In this paper we have considered the effect of vortex nucleation on
flux front propagation in type II superconductors. We have studied the
case of a constant applied magnetic field and compared the results of
numerical simulations with previous approaches to the problem
\cite{dorog,gilc,MOR-02}. While asymptotically we recover previous results,
a more accurate account of the vortex nucleation process leads to important
corrections for the scaling laws. This fact should be considered in the 
interpretation of experimental results.

\end{document}